\documentclass[aps, prb, superscriptaddress, reprint]{revtex4-1}
\usepackage{graphicx}
\usepackage{dcolumn}
\usepackage{bm}
\usepackage{amsmath}

\usepackage{color}
\usepackage{bm,graphicx,hyperref}
\hypersetup{%
  breaklinks = {true},
  citecolor = {blue},
  colorlinks = {true},
  linkcolor = {red},
}

\begin{document}

\title{Many-impurity phonon Casimir effect in atomic chains}

\author{Aleksandr Rodin}
\affiliation{Yale-NUS College, 16 College Avenue West, 138527, Singapore}
\affiliation{Centre for Advanced 2D
  Materials, National University of
  Singapore, 6 Science Drive 2, 117546, Singapore}

\date{\today}

\begin{abstract}

Phonon Casimir effect is the long-range interaction between impurities in condensed matter systems, mediated by vacuum fluctuations of the phonon field. For pairs of impurities, this interaction has been shown to follow a quasi-power law at zero-temperature and evolve into an exponentially decaying form as the temperature is increased. This work introduces an approach to deal with systems of more than two impurities, both at zero and finite temperatures.

\end{abstract}

\maketitle

\section{Introduction}
\label{sec:Introduction}

Phonons in solid state systems are typically discussed either in the context of the material heat capacity or as a scattering mechanism for charge carriers. In the latter case, they are generally seen as a nuisance since scattering suppresses electronic transport. At the same time, it is precisely this electron-phonon coupling that leads to the formation of Cooper pairs and, by extension, superconductivity.~\cite{Bardeen1957tos}

The phonons that bind Cooper pairs arise due to the lattice distortion caused by moving electrons. However, particle propagation is not a requirement for phonon-mediated interaction. Because of the Heisenberg uncertainty principle, even at absolute zero, a phonon system contains vacuum fluctuations. When these virtual phonons scatter off impurities, they produce a force between these impurities.~\cite{Schecter2014pmc, Pavlov2018pmc, Pavlov2019} Since this coupling is rooted in the zero-point energy of the host's phonon modes, it has been dubbed the ``phonon Casimir effect" (PCE) to highlight its similarity to the original electromagnetic Casimir effect.~\cite{Casimir1948ota}

One should note that bulk-mediated coupling between impurities is not uncommon in condensed matter systems. For example, it has been shown that bulk electrons can give rise to attractive or repulsive interaction.~\cite{Shytov2009lri, Lebohec2014art, Agarwal2017lre, Agarwal2019pas, Rodin2019bmi} Because this interaction is closely related to the Friedel oscillations, it can change sign with the impurity-impurity separation. What makes PCE different is that its sign is determined by the relative mass of the impurities and the host system's atoms. Another key difference between the two is the importance of temperature. Varying the temperature in the electron-mediated case is not expected to have a dramatic effect because the number of electrons in a metal is determined by the chemical potential. The number of phonons, on the other hand, strongly depends on the temperature of the system making the interaction much more temperature-sensitive.~\cite{Schecter2014pmc, Pavlov2019}

Earlier work on the subject~\cite{Schecter2014pmc, Pavlov2018pmc, Pavlov2019} dealt with the interaction between pairs of impurities. The aim of this study is to explore PCE for an arbitrary impurity configuration at finite temperatures using the path integral formalism. Because of the short-range nature of the phonon scattering, it is possible to solve the problem without resorting to the diagrammatic approach. To make the derivation as transparent as possible, the system considered in this work consists of a one-dimensional atomic chain of identical atoms. At the same time, the approach outlined here is general and can be adapted to other systems.

The model for a one-dimensional system with an arbitrary number of impurities is introduced in Sec.~\ref{sec:Model}. The illustration of the interaction effects is given in Section~\ref{sec:Impurity_Interaction}. Finally, the summary and conclusions can be found in Sec.~\ref{sec:Summary}.

\section{Model}
\label{sec:Model}

The one-dimensional phonon system consists of a periodic harmonic chain of atoms with mass $m$ connected by springs with the force constant $K$. Replacing some of the atoms with impurities gives the following Hamiltonian:
\begin{align}
	\hat{H} &= \sum_j \left[\frac{\hat{p}^2_j}{2m}+\frac{K}{2}\left(\hat{u}_j - \hat{u}_{j - 1}\right)^2\right] 
	\nonumber
	\\
	&+ \sum_l \frac{\hat{p}_l^2}{2}\left(\frac{1}{M_l}-\frac{1}{m}\right)\,.
	\label{eqn:Hamiltonian_Real_Space}
\end{align}
Here, the index $j$ runs over all the atoms in the chain, $l$ corresponds to the atoms replaced by the impurities, and $M_l$ is the impurity mass. Note that $M_l > m$ can be achieved by either swapping out a chain atom for an atom of a heavier mass or by introducing an adatom which binds to the chain so that the combined mass of the host and the adsorbate is $M_l$. In the latter case, all internal dynamics of the host atom-impurity subsystem are ignored by approximating it as a single composite object in order to keep the problem as simple as possible. For the same reason, only oscillations in the direction of the chain are considered, leading to a single phonon branch. 

Taking the Fourier transform of Eq.~\eqref{eqn:Hamiltonian_Real_Space} and performing the standard harmonic oscillator substitution~\cite{Bruus2002} with $\hbar \rightarrow 1$ yields the second-quantized Hamiltonian
\begin{align}
	\hat{H} & =  
	 \sum_{qq'l}
	\frac{\alpha_l}{4N} 
	\sqrt{\Omega_q}\sqrt{\Omega_{q'}}
	\left(b^\dagger_{-q} - b_q\right)
	\left(b^\dagger_{q'} - b_{-q'}\right)
	e^{ir_l\left(q - q'\right)}
	\nonumber
	\\
	&+\sum_q \Omega_q \left(b_q^\dagger b_q + \frac{1}{2}\right)
		\,,
	\label{eqn:Hamiltonian_Second_Quantization}
\end{align}
where $N$ is the number of atoms in the chain, $\alpha_l = 1-m/M_l$, $\Omega_q^2 = \Omega^2\sin^2\left(qa / 2\right)$, $\Omega = 2\sqrt{K/ m}$, and $a$ is the lattice constant.

A Hamiltonian expressed using the creation and annihilation operators is easily translatable into  the action as long as the operators are normal ordered. To establish the correct ordering in Eq.~\eqref{eqn:Hamiltonian_Second_Quantization}, one needs to expand the operator product in the first term and commute $b_q$ with $b^\dagger_{q'}$. Performing the commutation gives rise to a scalar quantity $- \sum_{q,l} \frac{\alpha_l}{4N} \Omega_q$, a correction to the zero-point energy of the impurity-free chain. This correction does not depend on the adatom positions and, therefore, plays no role in their interaction. For this reason, it will be temporarily neglected along with the constant $\sum_q \Omega_q / 2$ part of the second term.

Keeping only the terms with operators from Eq.~\eqref{eqn:Hamiltonian_Second_Quantization} gives the imaginary-time action
\begin{align}
	S &= \sum_{q\omega_n} \bar{\phi}_{q,\omega_n} \left(-i\omega_n + \Omega_q\right) \phi_{q,\omega_n}
	\label{eqn:S}
	\\
	&- \sum_{qq'\omega_nl}
	\begin{pmatrix}
		\bar{\phi}_{-q,-\omega_n} &\phi_{q,\omega_n}
	\end{pmatrix}
	\alpha_l
	\frac{Y_{q,l}Y^\dagger_{q',l}}{4N}
	\begin{pmatrix}
		\phi_{-q',-\omega_n}  \\ \bar{\phi}_{q',\omega_n} 
	\end{pmatrix}
	\nonumber
	\,,
\end{align}
where $Y_{q,l}^\dagger 
= 
\begin{pmatrix}
	1 & -1
\end{pmatrix}\sqrt{\Omega_q} e^{-ir_lq}$. This can be rewritten in a more symmetric fashion as
\begin{align}
	&S 
	=-\frac{1}{2} \sum_{qq'\omega_n}
	\begin{pmatrix}
		\bar{\phi}_{-q,-\omega_n} &\phi_{q,\omega_n}
	\end{pmatrix}
	\hat{\Gamma}_{qq'\omega_n}^{-1}
	\begin{pmatrix}
		\phi_{-q',-\omega_n} \\ \bar{\phi}_{q',\omega_n}
	\end{pmatrix}
	\label{eqn:S_Compact}
	\,,
	\\
	&\hat{\Gamma}_{qq'\omega_n}^{-1}  = \hat{G}^{-1}_{q\omega_n}
	\delta_{q,q'}
	+ 
	 \sum_l \frac{Y_{q,l} \alpha_l  Y_{q',l}^\dagger}{2N}\,,
	\label{eqn:Gamma_Inv}
	\\
	& \hat{G}^{-1}_{q\omega_n}=\begin{pmatrix}
		-i\omega_n - \Omega_q & 0
		\\
		0 & i\omega_n - \Omega_q
	\end{pmatrix}\,.
	\label{eqn:G_Inv}
\end{align}
The matrix $\hat{G}_{q\omega_n}$ is the two-frequency Green's function for a pristine chain, while $\hat{\Gamma}_{qq\omega_n}$ is the same for a chain with impurities. The factorized coupling term in Eq.~\eqref{eqn:Gamma_Inv} originating from the short-range scattering makes the inversion quite simple:
\begin{equation}
	\hat{\Gamma}_{\omega_n} = \hat{G}_{\omega_n} 
	- 
	\frac{1}{2N}
	\hat{G}_{\omega_n} \mathcal{Y}\alpha 
	\left(\underbrace{1+\frac{\mathcal{Y}^\dagger \hat{G}_{\omega_n} \mathcal{Y} \alpha}{2N}}_{\Delta_{\omega_n}}\right)^{-1}
	\mathcal{Y}^\dagger \hat{G}_{\omega_n}\,,
	\label{eqn:Gamma}
\end{equation}
where $\mathcal{Y} = \begin{pmatrix} Y_1 & Y_2 & \cdots \end{pmatrix}$, $Y_j$ is the column vector of $Y_{q,j}$, $\alpha$ is the a diagonal matrix of $\alpha_l$, and $\hat{G}_{\omega_n}$ is a block-diagonal matrix with entries given by the inverse of Eq.~\eqref{eqn:G_Inv}.

Exponentiating $-S$ and integrating over all fields yields the partition function $\mathcal{Z}$. While the integral itself is not problematic due the bilinear nature of $S$, one should take note of the structure of the action. Typically, for non-interacting systems, $S$ has the form of $\bar\Phi \mathbf{M} \Phi$, where $\Phi$ ($\bar\Phi$) is a vector of $\phi$ ($\bar\phi$) fields. This is because the Hamiltonian contains only creation-annihilation products. In this case, however, Eq.~\eqref{eqn:Hamiltonian_Second_Quantization} also includes $b^\dagger b^\dagger$ and $bb$ terms so that the action takes the $\begin{pmatrix}\bar\Phi & \Phi \end{pmatrix} \mathbf{M} \begin{pmatrix}\Phi \\ \bar \Phi \end{pmatrix}$ form. By performing a coordinate transformation, the integration fields can be changed from complex to twice as many real ones. Integrating over these new fields gives $\mathcal{Z} = \prod_{\omega_n}\det\left[-\beta\hat{\Gamma}_{\omega_n}^{-1}\right]^{-1/2}$. The power $1/2$ in the result is the consequence of the integration being performed over real fields. Naturally, the coordinate transformation also changes the matrix $\mathbf{M}$. However, the determinant is invariant under unitary transformations, hence one can use the determinant of the original matrix.

The free energy of the system can be obtained from the partition function using $F = -T \ln\mathcal{Z}$. The fact that the logarithm of the determinant equals the trace of the logarithm allows one to perform a series of simplifications to get
\begin{align}
	F &= \frac{1}{2}\sum_q\Omega_q + \frac{T}{2} \sum_{\omega_n} 
	\mathrm{tr}\, \ln
	\left(
	-\beta\hat{G}_{\omega_n}^{-1} \right)
	\nonumber
	\\
	&-\frac{1}{4N}\left(\sum_l \alpha_l\right)\sum_{q}\Omega_q+ \frac{T}{2} \sum_{\omega_n} 
	\mathrm{tr}\, \ln 
	\Delta_{\omega_n}
	\,.
	\label{eqn:F}
\end{align}
Note the reintroduction of the zero-point energy terms. The first row corresponds to the free energy of a pristine chain, composed of the vacuum energy and finite-$T$ excitations. The second row is the free energy due to the impurities. Explicitly,
\begin{equation}
	\Delta_{\omega_n} = 1
	+
	\begin{pmatrix}
		P_{11}\left(i\omega_n\right)
		&
		P_{12}\left(i\omega_n\right)
		&
		\cdots
		\\
		P_{21}\left(i\omega_n\right)
		&
		P_{22}\left(i\omega_n\right)
		&
		\cdots
		\\
		\vdots
		&
		\vdots
		&
		\ddots
	\end{pmatrix}\alpha \,,
	\label{eqn:Delta}
\end{equation}
where
\begin{align}
	&P_{jk}\left(i\omega_n\right)=Y_j^\dagger\frac{\hat{G}_{\omega_n}}{2N} Y_k 
	= -\frac{1}{N}
	\sum_q 
	e^{i\left(r_k-r_j\right)q}
	\frac{ \Omega_q^2}{\omega_n^2 + \Omega_q^2}
	\nonumber
	\\
	=&- \left[\delta_{k,j} - \sqrt{-\frac{\left(i\omega_n\right)^2}{\Omega^2-\left(i\omega_n\right)^2}}
	\right.
	\label{eqn:P}
	\\
	\times&
	\left.
	\left(1 - 2\frac{\left(i\omega_n\right)^2}{\Omega^2} - 2\sqrt{\frac{-\left(i\omega_n\right)^2}{\Omega^2}}\sqrt{1-\frac{\left(i\omega_n\right)^2}{\Omega^2}}\right)^{D_{jk}}\right]
	\nonumber
\end{align}
and $D_{jk} = \left|r_j - r_k\right| / a$ is number of lattice constants separating the impurities $j$ and $k$. From this, it is possible to write the impurity free energy $F_\mathrm{imp} = F_\mathrm{imp}^0 + F_I$ with
\begin{align}
	F_\mathrm{imp}^0 &= - \sum_{q,l}\frac{\alpha_l\Omega_q}{4N}+ \frac{T}{2} \sum_{\omega_n,l} 
	\ln \left[1 + \alpha_l P_{ll}\left(i\omega_n\right)\right]\,,
	\label{eqn:F_A_0}
	\\
	F_I
	&=
	\frac{T}{2} \sum_{\omega_n} 
	\ln
	\left|
	\begin{pmatrix}
		1+\alpha_1 P_{11}\left(i\omega_n\right)
		&
		\cdots
		\\
		\vdots
		&
		\ddots
	\end{pmatrix}^{-1}\Delta_{\omega_n}
	\right|
	\,,
	\label{eqn:F_I}
\end{align}
where the matrix multiplying $\Delta_{\omega_n}$ is diagonal with $1+\alpha_l P_{ll}\left(i\omega_n\right)$ as the elements. The quantity $F_\mathrm{imp}^0$ is the contribution to the free energy due to the impurities without the impurity-impurity interaction effects. The interaction portion of the energy is captured by the $F_I$ term.

To calculate any of the impurity energy terms, it is necessary to perform the frequency summation of the general form $\Sigma = \frac{T}{2}\sum_{\omega_n}f\left(i\omega_n/\Omega\right)$. Given that the singularities appear on the real axis, using the standard Matsubara techniques allows these sums to be written as
\begin{align}
	\Sigma 
	&=  \frac{\Omega}{2\pi}\int_{-\infty}^{\infty} n_B\left(\Omega z\right) \mathrm{Im}\left[f\left(z + i0\right)\right]dz\
	\nonumber
	\\
	&=  \frac{\Omega}{\pi}\int_{0}^{\infty}\left[ n_B\left(\Omega z\right) +\frac{1}{2}\right]\mathrm{Im}\left[f\left(z + i0\right)\right]dz\,.
	\label{eqn:Matsubara_Int}
\end{align}

It is important to note that the integrand in Eq.~\eqref{eqn:Matsubara_Int} becomes increasingly oscillatory as the number of impurities increases which can result in a loss of numerical precision. To assess the validity of a result obtained using Eq.~\eqref{eqn:Matsubara_Int}, one can compare it in the zero-temperature limit to the result calculated directly from Eq.~\eqref{eqn:F_I} by integrating over the Matsubara frequencies. Here, it is convenient to set $\omega_n\rightarrow \Omega \sinh \theta$ so that
\begin{align}
	F_I &=\frac{\Omega}{2\pi}\int_0^\infty d\theta\cosh\theta
	\ln
	\left|
	\begin{pmatrix}
		1+\alpha_1 P_{11}^\theta
		&
		\cdots
		\\
		\vdots
		&
		\ddots
	\end{pmatrix}^{-1}\Delta_{\theta}
	\right|\,,
	\nonumber
	\\
	P_{jk}^\theta &=- \left(\delta_{k,j} - e^{-2D_{jk}\theta}\tanh\theta \right)\,.
	\label{eqn:F_I_T0}
\end{align}

\section{Impurity Interaction}
\label{sec:Impurity_Interaction}

\subsection{Two Impurities}
\label{sec:Two_Impurities}

\begin{figure}
	\includegraphics[width = 3in]{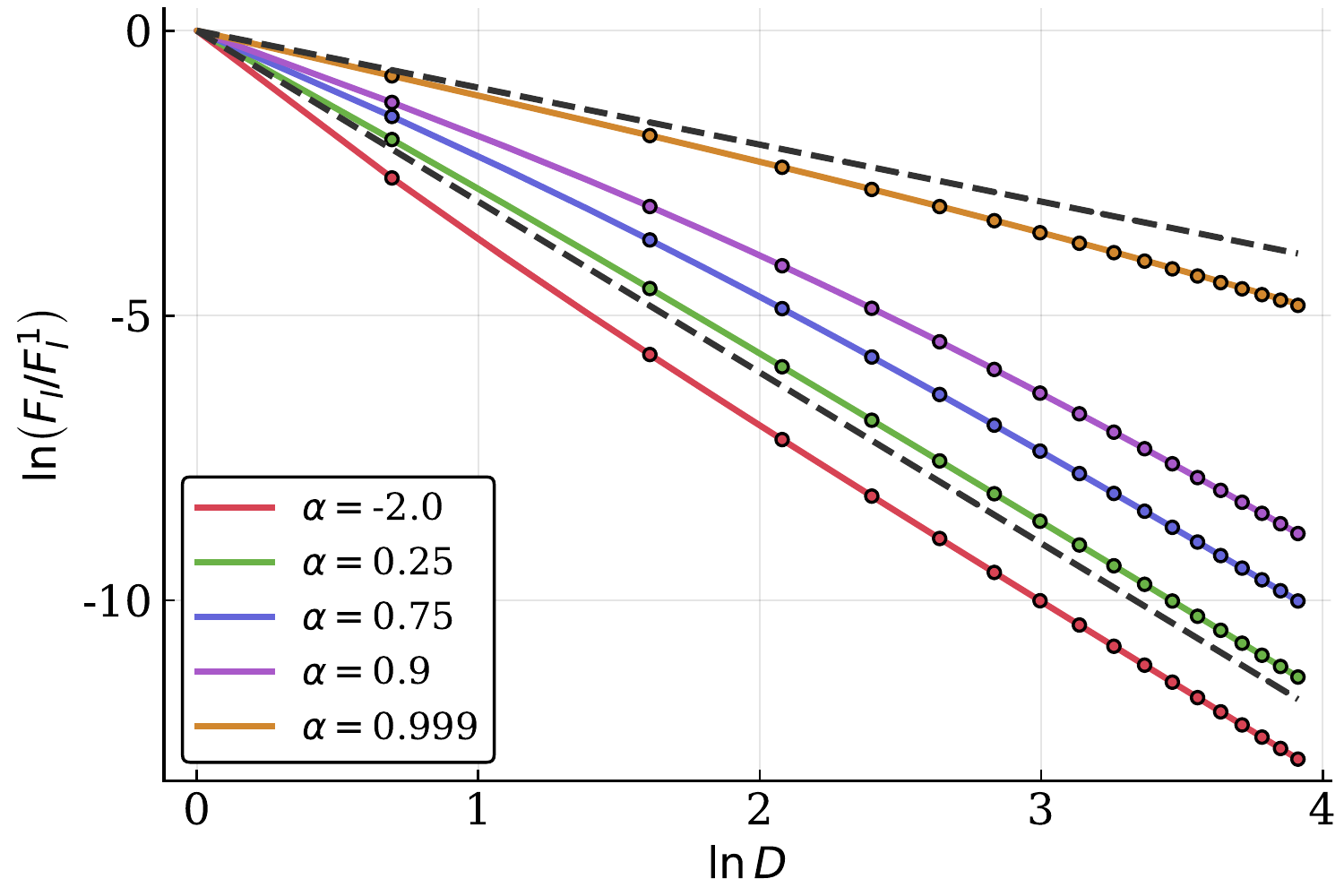}
	\\
	\includegraphics[width = 3in]{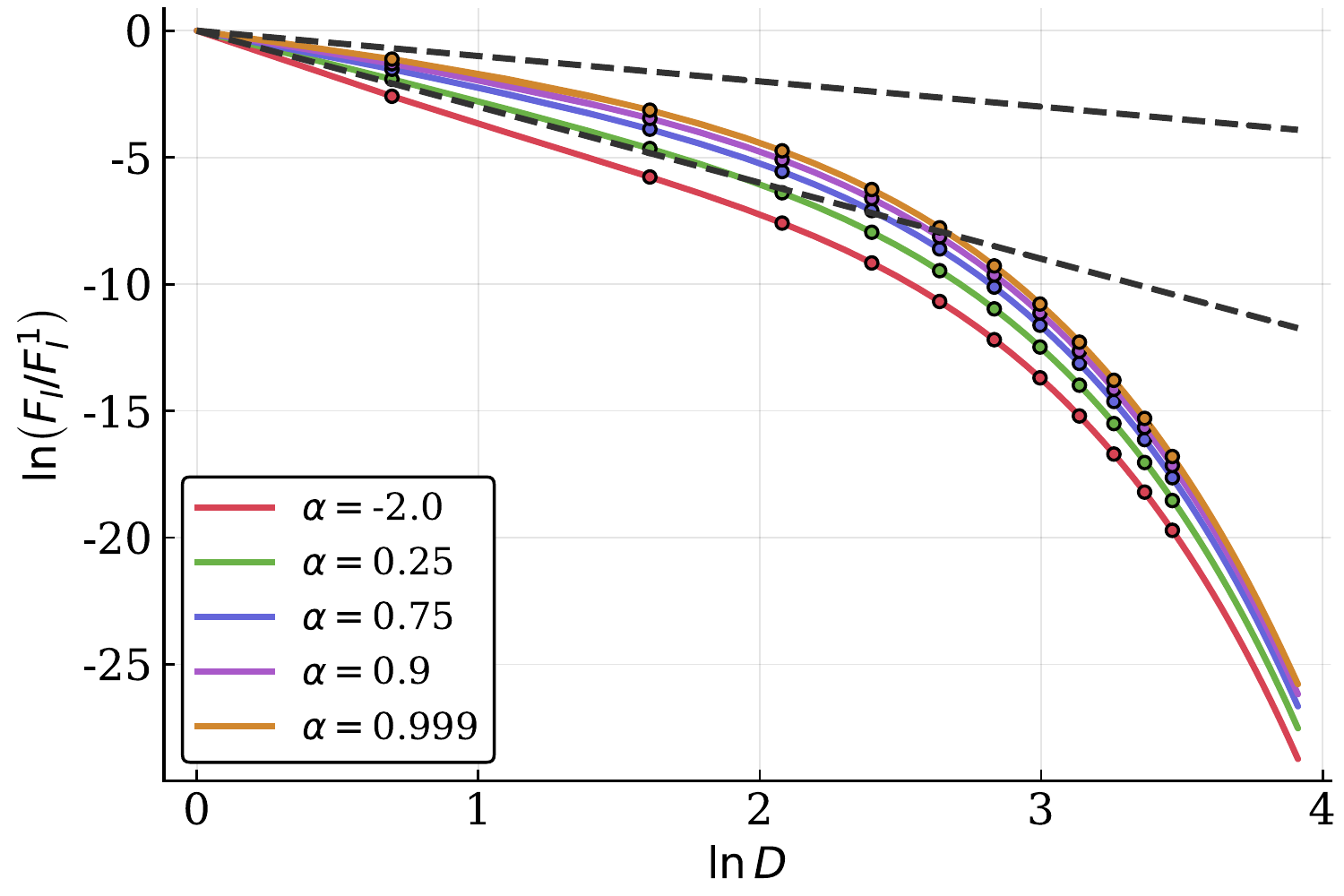}
	\caption{\textbf{Top:} $F_I$ at $T = 0$ calculated using Eq.~\eqref{eqn:F_I} for several values of $\alpha$. All the energies are normalized by dividing them by the corresponding $F_I^1$, the interaction of adjacent impurities. The dashed lines mark the $D^{-1}$ and $D^{-3}$ power laws. \textbf{Bottom:} Same as the top panel with $T = 0.02\Omega$. The points on the curves correspond to energies obtained from exact diagonalization for $N = 1000$. The bottom curves contain fewer points due to the loss of numerical precision for large $D$ when $F_\mathrm{exact}\left(N/2\right)$ is subtracted from $F_\mathrm{exact}\left(D\right)$.}
	\label{fig:Interaction}
\end{figure}

To validate the formula in Eq.~\eqref{eqn:F_I}, $F_I$'s for pairs of identical impurities with various $\alpha$'s are plotted in Fig.~\ref{fig:Interaction}. In the zero-temperature case (top panel of Fig.~\ref{fig:Interaction}), the interaction energy exhibits a power-law-like dependence on $D$ with the exponent evolving from $-1$ to $-3$ as the separation between the impurities increases. This transition is faster for lighter impurities, in agreement with Ref.~\onlinecite{Pavlov2018pmc}.

At finite temperatures, the dependence of $F_I$ on $D$ changes dramatically (bottom panel of Fig.~\ref{fig:Interaction}). It is immediately obvious that the curves are no longer restricted to the region between the two dashed power-law lines as the interaction decays exponentially with $D$.~\cite{Schecter2014pmc, Pavlov2019}

In addition to calculating $F_I$ from Eq.~\eqref{eqn:F_I}, the interaction is also computed using exact diagonalization with the results given by the points on top of the curves in Fig.~\ref{fig:Interaction}. This is done by finding the eigenstates of a harmonic chain of $N$ atoms for the impurity separations $D$ and $N / 2$. One then computes the free energy of the system for both cases
\begin{equation}
	F_\mathrm{exact} = \sum_j T\ln\left[1-e^{-\tilde\Omega_j / T}\right]+ \frac{\tilde\Omega_j}{2}\,,
\end{equation}
where $\tilde\Omega_j$ are the mode energies. Finally, the interaction energy is obtained by taking the difference between the two free energies.

\begin{figure}
	\includegraphics[width = 3in]{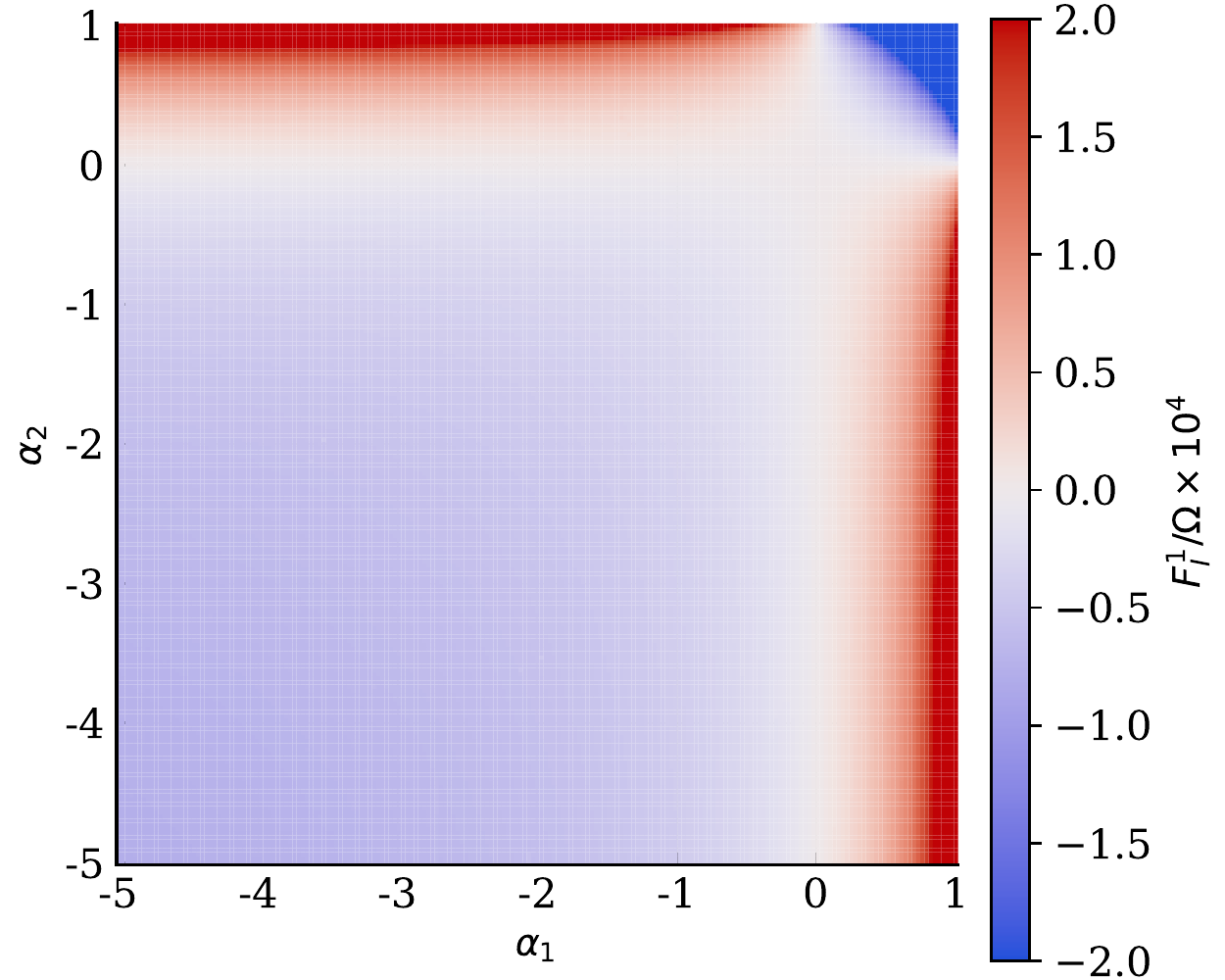}
	\caption{Zero-temperature $F_I$ at $D = 1$ for a two-impurity configuration. The sign of the interaction is given by $-\mathrm{sgn}\left(\alpha_1\alpha_2\right)$. The strongest interaction is between heavy impurities.}
	\label{fig:F_I^1}
\end{figure}

In the two-impurity case, zero-temperature $F_I$ is given by
\begin{align}
	F_I
	&=\frac{\Omega}{2\pi}\int_0^\infty d\theta \cosh \theta
	\label{eqn:F_I_2_T0}
	\\
	&\times
	\ln
	\left(
	1 - 
	\frac{\alpha_1\alpha_2\left[ 
	e^{-2D \theta}\tanh \theta
	\right]^2}{\left[1 -\alpha_1 \left(1 - \tanh \theta\right) \right]\left[1 -\alpha_2 \left(1-\tanh \theta\right) \right]}
	\right)
	\nonumber
	\,.
\end{align}
Due to the fact that $-\infty < \alpha\leq 1$, the denominator in the logarithm is positive. This means that the sign of the fraction and, therefore, the nature of the interaction is determined by the product $\alpha_1\alpha_2$, similar to an effect described in Ref.~\onlinecite{Schecter2014pmc} where changing the impurity parameters could be used to switch the interaction from attractive to repulsive. Since identical $\alpha$'s attract, opposite-sign $\alpha$'s will result in impurity repulsion. To illustrate this, $F_I$ at $D = 1$ for a range of $\alpha_1$ and $\alpha_2$ is plotted in Fig.~\ref{fig:F_I^1}.

Because the sign of the interaction depends on the sign of the product of $\alpha_1\alpha_2$, one may think of $\alpha$ as a type of ``charge". Unlike electric charges, however, here, like charges attract and opposite ones repel. This charge analogy is helpful for understanding the interaction of multi-impurity configurations. Equation~\eqref{eqn:F_I} shows that when the number of impurities is greater than two, the interaction energy is not given by the sum of two-impurity interactions. Nevertheless, as the first approximation, it is possible to treat the interaction as pairwise using the charge-like nature of the impurities in order to obtain the qualitative picture. One can then use Eq.~\eqref{eqn:F_I} to calculate the exact energy.

\subsection{Multiple Impurities}
\label{sec:Multiple_Impurities}

To appreciate the usefulness of the charge analogy, consider the following example. From Fig.~\ref{fig:F_I^1}, it is clear that at $T = 0$ the strongest interaction occurs between pairs of heavy impurities. At the same time, Fig.~\ref{fig:Interaction} shows that this interaction is very temperature-sensitive. With this in mind, let us position a negative-$\alpha$ impurity midway between two positive-$\alpha$ ones. For sufficiently low temperatures, the mutual attraction of the heavy impurities would cause them to approach each other despite the repulsion due to the light impurity. As the temperature is increased, the attractive term is weakened faster than the repulsive one causing the heavy impurities to separate.

\begin{figure}
	\includegraphics[width = 3in]{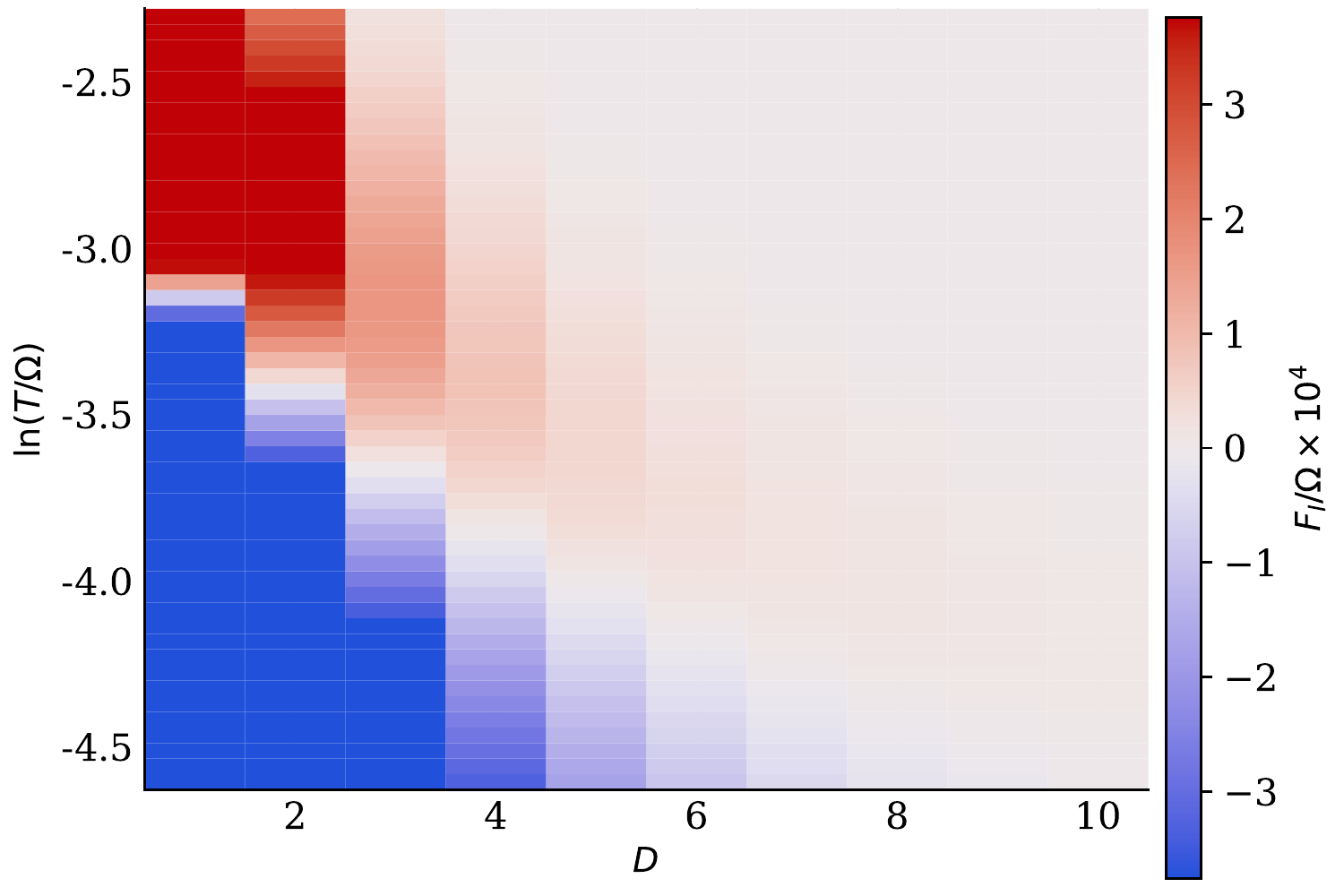}
	\caption{Interaction energy for heavy-light-heavy impurity configuration at different temperatures and separations. The heavy masses are $20m$ and the light one is $0.75m$. $D$ is the heavy-light separation. Varying the temperature at a fixed $D$ causes the interaction to change sign. For low-enough temperatures, the interaction transitions from attractive to repulsive with increasing $D$.}
	\label{fig:F_I_3_Imp}
\end{figure}

The interaction energy $F_I$ for this configuration is shown in Fig.~\ref{fig:F_I_3_Imp}. For this calculation, $D$ is the distance between the heavy impurities and the light one. One can see that for a given separation, increasing the temperature causes the interaction to transition from attractive to repulsive, as expected. In addition, at a fixed temperature, the interaction can be attractive at short distances and repulsive at large ones as the heavy-heavy attraction enters the exponentially decaying range.~\cite{Schecter2014pmc}

As an extension of the charge-like description, it is useful to address how these charges combine. If the impurities in the system correspond to chain atoms with adsorbates, $\alpha > 0$ and the adsorbates will tend to agglomerate and form continuous clusters. Using the multi-impurity formalism, it is possible to study how the attraction between these clusters depends on their size. Here, the analysis will be limited to a zero-$T$ case because of the numerical issues mentioned above.

Before performing the numerical computations, is helpful to establish a qualitative picture. By varying the impurity masses in the previous section, it was shown that the interaction energy is greater if the two chain segments are separated by heavier masses which lead to a reduced coupling between the segments. For infinitely-heavy impurities, therefore, increasing the cluster size should have no impact on the interaction energy because regardless of how many impurities the clusters contain, the two chain segments are completely isolated. For finite-mass impurities, on the other hand, increasing the cluster size results in a better separation between the segments. Naturally, the effect is expected to be more pronounced for lighter impurities since heavy impurities approach the almost-infinite mass regime even with small clusters.

\begin{figure}
	\includegraphics[width = 3in]{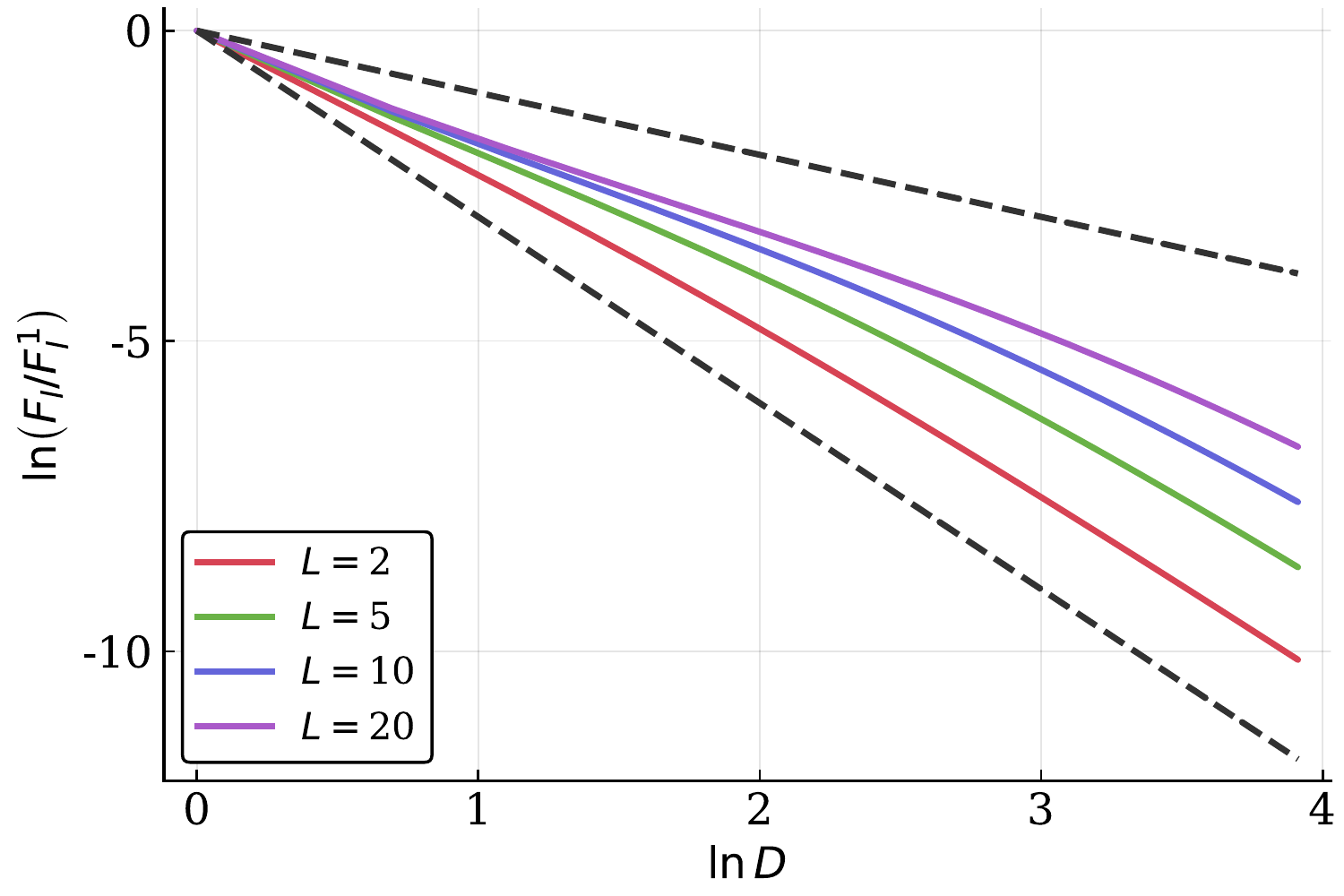}
	\includegraphics[width = 3in]{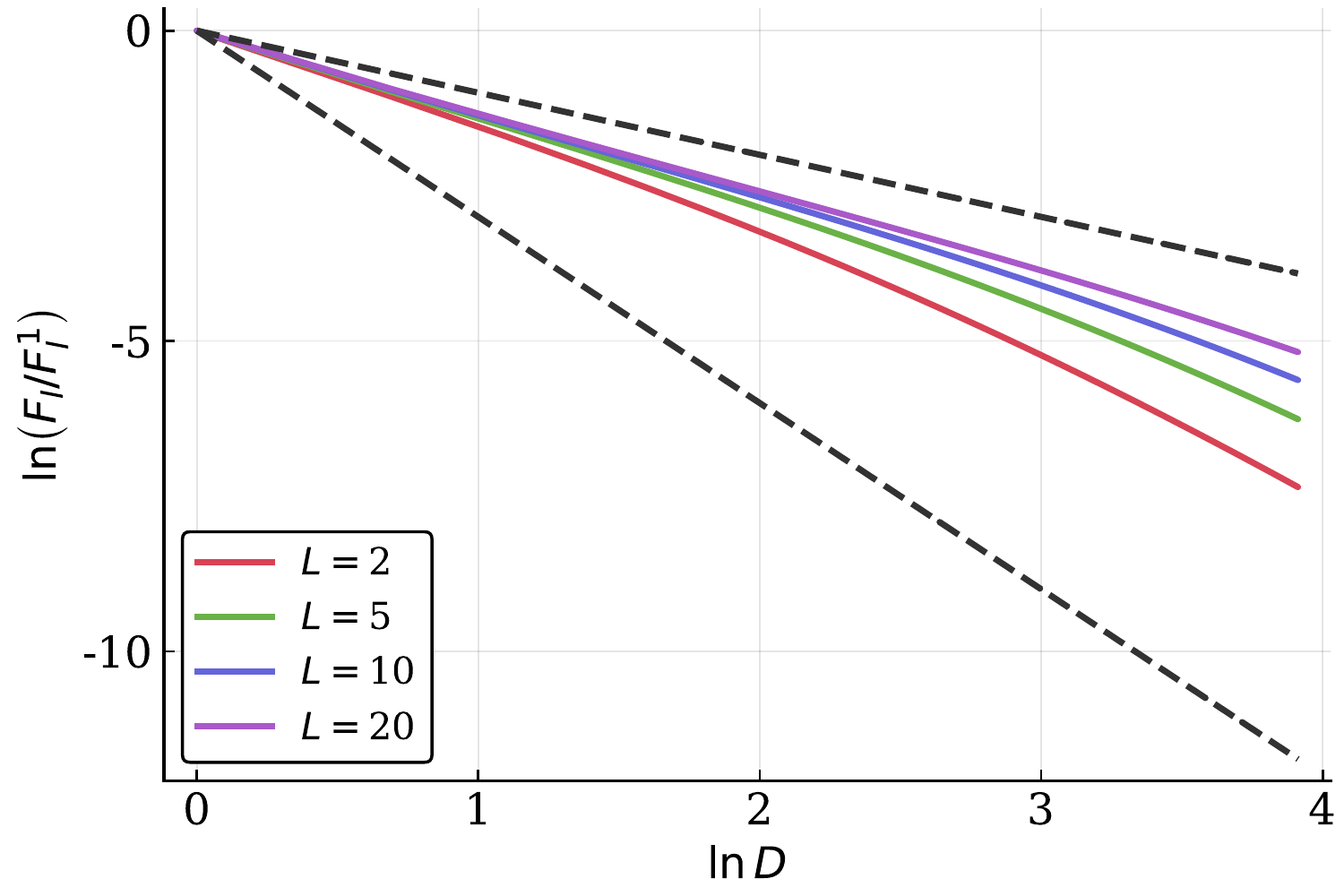}
	\caption{Interaction energy for different cluster lengths $L$ with $M = 1.5m$ (top) and $M= 15m$ (bottom). All energies are normalized by their corresponding $F_I^1$ which is the energy of the adjacent-cluster configuration.}
	\label{fig:Clusters}
\end{figure}

In order to calculate the two-cluster interaction energy using the formalism introduced in the previous section, one starts by first computing $F_I$ of the entire system. The cluster interaction is obtained by subtracting twice $F_I$ for individual clusters from the total energy. To illustrate how the sensitivity to the cluster size varies with $\alpha$, the interaction energy of cluster pairs is plotted in Fig.~\ref{fig:Clusters}.

As expected, increasing the cluster size for lighter impurities has a more dramatic effect on the cluster-cluster interaction than it does for heavier ones. For both impurity masses, increasing the cluster size gives diminishing returns, as can be seen by the reduced spacing between the curves at larger cluster lengths. Finally, the interaction energy demonstrates the quasi-power law dependence, approaching $D^{-3}$ for large separations, just as it does for individual masses.

\section{Summary}
\label{sec:Summary}

Interaction between impurities in PCE is not pairwise. Instead, the interaction energy needs to be computed for all impurities simultaneously. In this work, a non-diagrammatic solution for the interaction energy of a multi-impurity configuration was developed. This formula was validated by comparing its results with exact diagonalization. The proposed approach makes it possible to use minimization algorithms to find the minimum-energy impurity arrangement. In addition, this method is not limited to one-dimensional chains and can, with modification, be adapted to higher dimensions or multi-atomic lattices.

The numerical calculations were performed using Julia programming language.~\cite{Bezanson2017} The code is available at \href{url}{https://github.com/rodin-physics/1D-phonon-casimir}. The author acknowledges the National Research Foundation, Prime Minister Office, Singapore, under its Medium Sized Centre Programme and the support by Yale-NUS College (through grant number R-607-265-380-121).

\bibliography{Phononic_Casimir_1D}

\end{document}